\documentclass[lettersize,journal]{IEEEtran}
\usepackage{amsmath,amsfonts}
\usepackage{algorithmic}
\usepackage{algorithm}
\usepackage{array}
\usepackage[caption=false,font=normalsize,labelfont=sf,textfont=sf]{subfig}
\usepackage{textcomp}
\usepackage{stfloats}
\usepackage{url}
\usepackage{verbatim}
\usepackage{graphicx}
\usepackage{cite}
\usepackage{pifont}%
\newcommand{\cmark}{\ding{51}}%
\newcommand{\xmark}{\ding{55}}%

\begin{document}

\title{Optimal Operating Strategy for PV-BESS Households: Balancing Self-Consumption and Self-Sufficiency}
\author{Jun Wook Heo,
    Raja Jurdak,~\IEEEmembership{Senior Member,~IEEE}, Sara Khalifa
\thanks{Jun Wook Heo is with the School of Computer Science, Queensland University of Technology, Brisbane, QLD 4000, Australia (e-mail:junwook.heo@hdr.qut.edu.au}
\thanks{Raja Jurdak is with the School of Computer Science, Queensland University of Technology, Brisbane, QLD 4000, Australia (e-mail:r.jurdak@qut.edu.au)}

\thanks{Sara Khalifa is with the School of Information Systems, Queensland University of Technology, Brisbane, QLD 4000, Australia (e-mail:sara.khalifa@qut.edu.au)}
}

\markboth{Journal of \LaTeX\ Class Files,~Vol.~14, No.~8, August~2021}%
{Shell \MakeLowercase{\textit{et al.}}: A Sample Article Using IEEEtran.cls for IEEE Journals}

\IEEEpubid{0000--0000/00\$00.00~\copyright~2021 IEEE}

\maketitle

\begin{abstract}
High penetration of Photovoltaic (PV) generation and Battery Energy Storage System (BESS) in individual households increases the demand for solutions to determine the optimal PV generation power and the capacity of BESS. Self-consumption and self-sufficiency are essential for optimising the operation of PV-BESS systems in households, aiming to minimise power import from and export to the main grid. However, self-consumption and self-sufficiency are not independent; they share a linear relationship. This paper demonstrates this relationship and proposes an optimal operating strategy that considers power generation and consumption profiles to maximise self-consumption and self-sufficiency in households equipped with a PV-BESS. We classify self-consumption and self-sufficiency patterns into four categories based on the ratio of self-sufficiency to self-consumption for each household and determine the optimal PV generation and BESS capacities using both a mathematical calculation and this ratio. These optimal operation values for each category are then simulated using Model Predictive Control (MPC) and Reinforcement Learning (RL)-based battery charging and discharging scheduling models. The results show that the ratio between self-consumption and self-sufficiency is a useful metric for determining the optimal capacity of PV-BESS systems to maximise the local utilisation of PV-generated power.

\end{abstract}


\section{Introduction}
\IEEEPARstart{C}{ontemporary} environmental issues arising from conventional electricity generation have increased interest in renewable energy sources. Solar power, one of the fastest growing energy alternatives, is commonly harnessed through rooftop Photovoltaic (PV) panels in households \cite{kelepouris2021impact}. However, the imbalance between household electricity generation and consumption presents a challenge in both maximising the renewable energy utilisation and maintaining the safe operation of the smart grid \cite{chen2022transforming}. Energy storage systems, including Battery Energy Storage System (BESS), flywheel and hydrogen storage, are essential for maximising renewable energy utilisation and improving the grid resiliency. BESS, widely used at the household level, stores surplus power generated by PV systems and provides it during the peak demand periods, enhancing Self-Consumption (SC) and Self-Sufficiency (SS) \cite{bagalini2019solar,kelepouris2021impact,suthar2019cost}.

SC is the ratio of self-consumed PV electricity to total electricity generation from PV while SS is the ratio of self-consumed PV electricity to the total electricity demand \cite{bertsch2017drives,kelepouris2021impact}. Maximising SC and SS enhances grid independence by minimising power imports from and exports to the main grid \cite{nyholm2016solar}. For instance, to maximise SC, the power generated by PV should be consumed locally without exporting surplus energy to the main grid, which also reduces the total power imports from the main grid. On the other hand, SS affects the amount of imported and exported power. If the power generated by PV is insufficient to meet consumption needs, the potential reduction in import and export power through SC optimisation is limited. The optimal sizing of PV-BESS systems can impact the power injected into the main grid, thereby influencing overall energy flows \cite{ciocia2021self}. Optimising SC and SS in PV-BESS regulation enables smoother power import and export, effectively reducing the fluctuation of grid power \cite{li2013battery,yang2020fluctuation}. Therefore, optimising the size of PV and BESS should take into account both SC and SS simultaneously.

There are many studies on optimising the capacity of PV-BESS systems. For instance, several optimisation solutions for PV-BESS systems are proposed based on SC \cite{casu2020designing,blasuttigh2023optimal,duman2022optimal}, whereas an alternative approach focuses on SS \cite{korjani2022online}. Meanwhile, some studies consider both SC and SS for PV-BESS optimisations based on real dataset \cite{bertsch2017drives,nyholm2016solar,hassan2021self,hassan2022analysis,simoiu2021optimising}. The authors in  \cite{bertsch2017drives} identify the most profitable sizes of PV and storage systems and explores the levels of SC and SS while Array-go-Load ratio (ALR) is introduced to analyses SC and SS in Swedish households \cite{nyholm2016solar}. Additionally, the impact of an electric vehicle on the levels of SC and SS in households with a PV-BESS system is explored \cite{gudmunds2020self}. 

\IEEEpubidadjcol

However, existing solutions have not fully explored the relationship between SC and SS to determine the optimal capacity of PV and battery systems. SC and SS are not independent; rather they follow a linear relationship. The relationship between SC and SS is influenced not by the optimisation of BESS charging and discharging scheduling but by the amount of generation and consumption in households. Therefore, PV-BESS systems optimisation should account for this relationship to jointly maximise SC and SS.

This paper provides the mathematical definition of the relationship between SC and SS and introduces a strategy to determine the optimal capacity of PV-BESS systems based on this relationship. We present an optimal operation strategy to determine the capacities of  PV-BESS systems in a single household within a smart grid, considering the relationship between SC and SS. The proposed approach is evaluated using mathematical modelling and reinforcement learning solutions to maximise both SC and SS.
The key contributions of this paper are summarised below:
\begin{itemize}
    \item This paper mathematically defines the relationship factor between SC and SS, and conducts a comprehensive analysis using the Ausgrid dataset, which contains PV generation and consumption in households with PV-BESS systems.
    \item We proposed an analysis model, the Lightweight and Instantaneous Analysis for BESS scheduling (LITE-BESS), to calculate SC and SS from power generation and consumption profile in a household dataset.
    \item We classify households into four groups based on the relationship factor between SS and SC, utilising LITE-BESS to analysis power generation and consumption for a single household based on SS and SC. LITE-BESS employs simple strategies for scheduling BESS charging and discharging without involving complex calculations.
    \item Various BESS management models, including MPC and several RL approaches, are implemented and used to evaluate the identified household groups.
    \item We propose a scheme using the ratio of SS to SC for determining the optimal capacity of PV-BESS systems for a single household and validate it through simulations using MPC and RL models. The results are compared with those obtained from a metathetical solution and indicate that the ratio between SC and SS is a crucial factor in determining the capacities of PV-BESS systems.
\end{itemize}

\section{Related Work}
This section provides an overview of existing PV-BESS optimisation solutions for PV-BESS systems that utilise SC and SS. It then moves on to a detailed discussion of optimisation methods such as MPC and RL, which are employed to evaluate the performance of PV-BESS systems under various operating conditions.

\subsection{PV-BESS Optimisation}
For determining the optimal capacities for a home energy management system with PV-BESS, SC and SS are useful factors for maximising the use of locally generated power, while minimising both import of power from and export of power to the main grid. A recommendation tool for the optimal sizing of renewable energy source systems for nearly Zero Energy Buildings (nZEB) is proposed to maximise the building's SC while minimising the required investment \cite{casu2020designing}. The size of a condominium functioning as a Jointly Acting Renewable Self-Consumers (JARSC) community is investigated to improve SC, using a case study located in Milan, Italy \cite{blasuttigh2023optimal}. The optimal sizing of PV-BESS systems for households considers day-ahead load scheduling for demand response and SC \cite{duman2022optimal}. In contrast, an online energy management tool for sizing PV-BESS systems is presented based on their SS \cite{korjani2022online}. However, these solutions focus solely on SC or SS aspects individually without considering both SC and SS together.

On the other hand, a simulation model to identify the most profitable sizes of PV-BESS systems is presented, considering SC and SS, and comparing and contrasting Germany and Ireland to account for regulatory and geographical differences \cite{bertsch2017drives}.
PV-BESs systems in Swedish households are investigated, considering both SC and SS. ARL, which is the ratio of the PV capacity to the average annual electric load of a household, is introduced for different categories of single-family dwellings in Sweden \cite{nyholm2016solar}.  The impact of a supercapacitor module on PV systems to enhance SC and SS is evaluated under various supercapacitor capacities \cite{hassan2021self,hassan2022analysis}. The authors in \cite{simoiu2021optimising} formulate optimisation problems using criteria such as SC, SS and net present values (NPV) to determine the optimal PV size. However, as shown in Table \ref{tbl_comparison}, these approaches evaluate PV-BESS systems by considering SC and SS independently. Since SC and SS have a linear relationship, it is crucial to consider both simultaneously. In this paper, we demonstrate the linear relationship between SC and SS and validate it using various BESS management system models, including MPC and RL models.

\begin{table}[!ht]
\centering
\caption{Comparison of Existing Solutions.}
\begin{tabular}{|l|c|c|c|}
\hline
Solutions                   & SC & SS & $\rho = \frac{SS}{SC}$ \\[1mm] \hline
\cite{casu2020designing}    & \cmark  & \xmark & \xmark  \\ \hline
\cite{blasuttigh2023optimal}& \cmark  & \xmark & \xmark  \\ \hline
\cite{duman2022optimal}     & \cmark  & \xmark & \xmark  \\ \hline
\cite{korjani2022online}    & \xmark  & \cmark & \xmark  \\ \hline
\cite{bertsch2017drives}    & \cmark  & \cmark & \xmark  \\ \hline
\cite{nyholm2016solar}      & \cmark  & \cmark & \xmark  \\ \hline
\cite{hassan2021self}       & \cmark  & \cmark & \xmark  \\ \hline
\cite{hassan2022analysis}   & \cmark  & \cmark & \xmark  \\ \hline
\cite{simoiu2021optimising} & \cmark  & \cmark & \xmark  \\ \hline
\textbf{LITE-BESS}    & \cmark  & \cmark & \cmark  \\ \hline
\hline
\end{tabular}
\label{tbl_comparison}
\end{table}

\subsection{MPC}
The MPC approach is commonly used for optimising BESS charging and discharging schedules in the smart grid. \cite{mbungu2020dynamic,amani2023data,trigkas2021virtual,zhang2020mpc}. MPC predicts future behaviour of the system by measuring or estimating current state of the system, hypothetical future inputs and constraints \cite{kouvaritakis2016model}. For a linear discrete-time state-space model, the future system state can be estimated based on the current system state and input as follows:
\begin{equation}
    x_{k+1} = Ax_k + Bu_k
\end{equation}
Where A and B are the system and input matrices, respectively. The output is defined as follows:

\begin{equation}
    y_{k} = Cx_k
\end{equation}
Where C is output matrices. The quadratic objective function can be represented with the distance between the predicted output $\hat{y}$ and a given reference $y$ for a horizon N, so MPC minimises a user-defined cost function $J$ as follows:

\begin{equation}
    J(k) = \sum_{i=1}^{N} [y(k+i|k) - \hat{y}(k+i|k)]^2 + \lambda \sum_{i=0}^{N-1} [\Delta u(k+i|k]^2
\end{equation}
Where the control effort penalization term $\Delta u(k) = u(k) - u(k-1)$ and $\lambda$ is a weighting factor for the control effort penalization term \cite{faedo2017optimal,schwenzer2021review}.

\subsection{MDP}
In the discounted Markov Decision Process (MDP) $(X , A, \gamma, P, r)$, an action $a_t \in A$ at time $t$ is chosen based on the policy $\pi_\theta(a|s_t)$, given its current state $s_t \in X$. The reward is represented as $r(s_t, a_t)$ and the state transitions to the next state $s_{t+1}$ according to the probability $P(s_{t+1}|s_t, a_t)$. The objective of the agent is to maximise the cumulative return $J(\theta) = \mathbb{E}_{\pi} [R_t] = \mathbb{E}_{\pi} \left[ \sum_{i=0}^{\infty} \gamma^i r(s_{t+i}, a_{t+i}) \right]$, where $\mathbb{E}$ indicates the expectation \cite{wu2017scalable,arulkumaran2017deep}.

\subsection{DQN}
The Deep Q-network (DQN) uses a deep neural network to approximate the optimal action values in a given sate. The optimal Q-function can be represented in combination of the current reward and the future reward as follows:
\begin{equation}
    Q^* (s, a) = \mathbb{E}[r + \gamma \max_{a'} Q^\star (s', a')|s, a]
\end{equation}
Where $\gamma$ is the discount factor for future rewards. Using Q-table containing each combination of state and action is impractical, so DQN trains the network estimating the Q-values, i.e., $Q(s, a; \theta) \approx Q^*(s,a)$. Thus, the loss at iteration i is defined as follows:
\begin{equation}
    L_i (\theta_i) = \mathbb{E} [(y_i - Q(s, a; \theta_i)^2]
\end{equation}
Where the temporal difference target, $y_i = r + \gamma \max_{a'} Q(s', a'; \theta_{i-1})$. To stabilise neural network updates, DQN employs an experience replay buffer. Transition data are added to the buffer and samples from the buffer are used to train DQN \cite{mnih2013playing}. \cite{bui2019double,wang2023dqn,doan2023deep,kim2023optimize} introduce DQN models for the optimal operation of energy storage systems.

\subsection{A2C}
The Advantage Actor-Critic (A2C) method consists of an actor network and a critic network. The actor network selects actions based on the current state of the environment while the critic network estimates the expected return for a given state under the current policy. The action value function $Q(s,a)$ can be expressed in terms of the current reward and the estimated value of the future state as follows:
\begin{equation}
    Q(s, a) = r + \gamma V(s^\prime)
\end{equation}
The advantage is defined as follows:
\begin{equation}
    A(s_t, a_t) = Q(s_t, a_t) - V(s_t) = r + \gamma V(s_{t+1}) - V(s_t) 
\end{equation}
The policy gradient of A2C is given by:
\begin{equation}
    \Delta_\theta J_\pi(\theta) = \mathbb{E}_{\pi_\theta}[\Delta_\theta \, log\pi_\theta(a_t|s_t) \, A(s_t, a_t)]
\end{equation}
The critic network updates its parameters by minimising $J_Q(\varphi)$, which is defined as follows \cite{chen2020intelligent,kang2024reinforcement,hua2021data,selim2022optimal,qin2021privacy}:
\begin{equation}
    J_Q(\varphi) =(r + \gamma V(s^\prime) - V(s_t))^2
\end{equation}

\subsection{PPO}
Proximal Policy Optimisation (PPO) is also based on the actor-critic architecture, clipping the policy gradient. To optimise policies, sampling data from the policy is used to train for several epochs \cite{schulman2017proximal,huang2020deep,li2022ppo,meng2023proximal,kang2023multi,anwar2022proximal}. The objective of PPO is defined as follows:
\begin{equation}
    L^{CLIP}(\theta) = \mathbb{E}_t[min(r_t(\theta) \, A_t, \, clip(r_t(\theta), 1 - \epsilon, 1 + \epsilon) \, A_t)]
\end{equation}
Where the probability ratio $r_t(\theta) = \frac{\pi_\theta(a_t|s_t)}{\pi_{\theta_{old}}(a_t|s_t)}$.

\section{Methodology}
In this section, we introduce LITE-BESS based on straightforward strategies for managing BESS charging and discharging scheduling, which operates ultra-fast. Then, $\rho$, the ratio of SS to SC, is mathematically defined, and based on $\rho$ we categorise households into four groups using the Ausgrid dataset \cite{ratnam2017residential}. The data was sourced from 300 randomly selected solar customers in Ausgrid, covering the period from July 1, 2010, to June 30, 2013. The dataset primarily includes PV generation and consumption power profiles, measured every 30 minutes for each household. The optimal operation capacities of PV-BESS systems are calculated mathematically.

\subsection{Analyser Model}
\begin{figure}[!ht]
\centering
\includegraphics[width=0.45\textwidth]{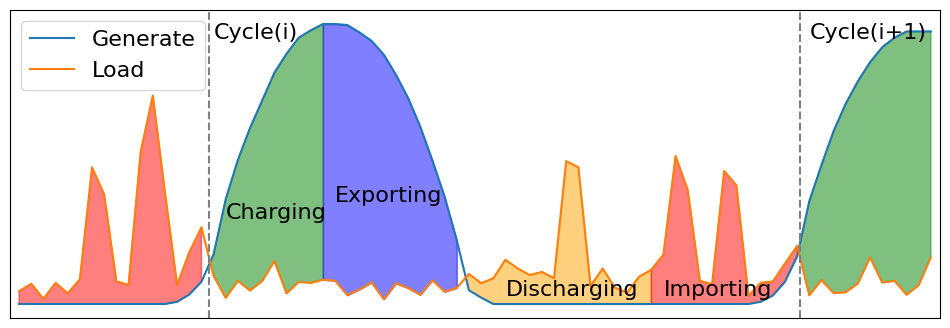}
\caption{LITE-BESS.}
\label{fig_analizer}
\end{figure}

For an effective and efficient analysis of SC and SS, we define three strategies for LITE-BESS, as shown in Fig.~\ref{fig_analizer}:
\begin{itemize}
\item A cycle is defined as the period between when power generation exceeds consumption and the next occurrence of this condition. The proposed model operates based on this cycle.
\item If the generated power exceeds the consumed power, the model first charges the battery unless it is already full. If the battery is full, the surplus power is exported to the main grid.
\item If the consumed power exceeds the generated power, the model first discharges the battery. However, if the battery is empty, the insufficient power is imported from the main grid.
\end{itemize}

Based on these strategies, the surplus power in the $i-th$ cycle is defined as follows:
\begin{equation}
\begin{split}
    P_{surplus}^i = \sum_{t=0}^{T} (P_{gen}^i(t)-P_{load}^i(t)) \\ 
    \text{, if } P_{gen}^i(t)>P_{load}^i(t) 
\end{split}
\end{equation}
Where $T$ is the duration of the $i-th$ cycle. $P_{gen}^i(t)$ and $P_{load}^i(t)$ represent the total generated power and consumed power during the $i-th$ cycle, respectively.

Likewise, the deficit power in the $i-th$ cycle is defined as follows:
\begin{equation}
\begin{split}
    P_{deficit}^i = \sum_{t=0}^{T} (P_{load}^i(t)-P_{gen}^i(t)) \\ 
    \text{, if } P_{load}^i(t)>P_{gen}^i(t) 
\end{split}
\end{equation}

The total amount power for charging and discharging during the $i-th$ cycle can be represented as follows:
\begin{equation}
    P_{charging}^i(t) = min(SOC_{max} - SOC_{init}^i, P_{surplus}^i)
\label{eq.pcharging}
\end{equation}

\begin{equation}
    P_{discharging}^i(t) = min(SOC_{state}^i - SOC_{min}, P_{deficit}^i)
\label{eq.pdischarging}
\end{equation}

Where $SOC_{init}^i$ and $SOC_{state}^i$ represent the initial state of charge and the battery state in the $i-th$ cycle, respectively. From the Eq.~\ref{eq.pcharging} and Eq.~\ref{eq.pdischarging}, $(i+1)-th$ initial state of charge can be calculated as follows:

\begin{equation}
    SOC_{state}^{i+1} = SOC_{init}^i + P_{charging}^i - P_{discharging}^i 
\end{equation}

During the $i-th$ cycle, the total amount of imported and exported power from/to the main grid are defined as follows:

\begin{equation}
    P_{import}^{i} = P_{deficit}^i - P_{discharging}^i
\end{equation}

\begin{equation}
    P_{export}^{i} = P_{surplus}^i - P_{charging}^i
\end{equation}

As the analyser model is very simple and does not include complex calculation for optimising the charging and discharging schedule, it takes about 20 milliseconds to calculate SC and SS for household data measured every half hours over a year.

\subsection{Categorising Household Types}
SC and SS can be calculated as follows:

\begin{equation}
    SC = 1 - \frac{\sum_{i=0}^{N-1} P_{export}^i}{\sum_{i=0}^{N-1} P_{gen}^i}
\end{equation}

\begin{equation}
    SS = 1 - \frac{\sum_{i=0}^{N-1} P_{import}^i}{\sum_{i=0}^{N-1} P_{load}^i}
\end{equation}
Where N is the total number of cycles in the entire period. The amount of power remaining in BESS is negligible if the battery capacity is small compared to the total generation and consumption power in a household during the operation period. Additionally, the sum of the total generation and import power equals the sum of the total consumption and export power in a single household. Thus, we define $\rho$, which represents the ratio of SS to SC as follows:

\begin{equation}
    \rho = \frac{SS}{SC} \simeq \frac{\sum_{i=0}^{N-1}P_{gen}^{i}}{\sum_{i=0}^{N-1}P_{load}^{i}}
\label{eq_rho}
\end{equation}

Based on the LITE-BESS, we analyse SC and SS for each household of Ausgrid dataset and categorise them into four groups as follows:

\begin{itemize}
    \item Type 1 : $ 0 \leq \rho < 0.4 $ (e.g., Household \#25)
    \item Type 2 : $ 0.4 \leq \rho < 0.7 $ (e.g., Household \#93)
    \item Type 3 :  $ 0.7 \leq \rho < 1.0 $ (e.g., Household \#48)
    \item Type 4 : $ 1.0 \leq \rho $ (e.g., Household \#75)
\end{itemize}

\begin{figure}[!ht]
\centering
\includegraphics[width=0.45\textwidth]{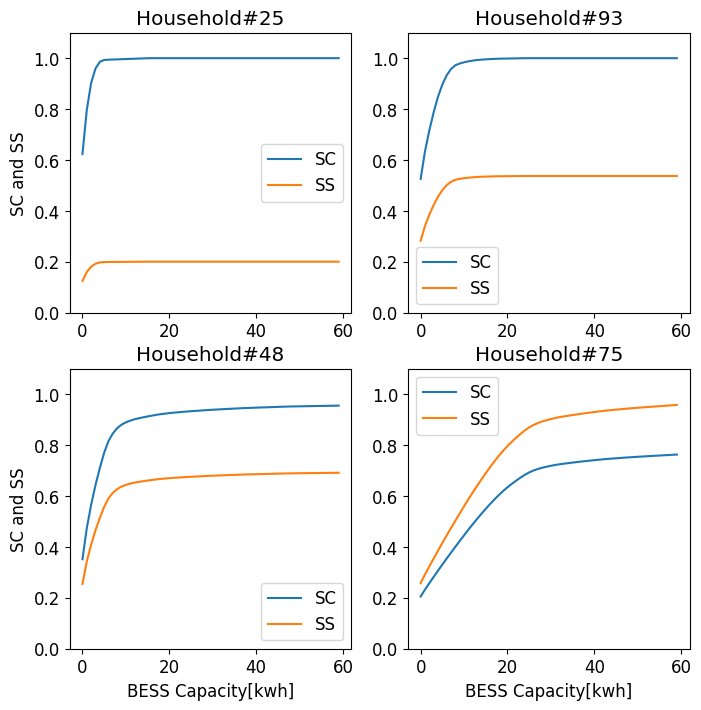}
\caption{$SC and SS$ for Different Household Types.}
\label{fig_household_category}
\end{figure}

Fig.~\ref{fig_household_category} illustrates SC and SS in Households \#25, \#93, \#48 and \#75 according to the capacity of the BESS. Due to the insufficiency of generation power, the SS of Type 1 and Type 2 remains low even as the BESS capacity increases, whereas SC quickly reaches a high level at lower BESS capacities. On the other hand, the SC of Type3 and Type 4 is lower compared to Type1 and Type2. The SC and SS of Type 3 and Type 4 increase rapidly at first and then rise more slowly as the BESS capacity increases. To sum up, as $\rho$ increases, SC decreases, while SS and the optimal BESS capacity increases. This indicates that installing additional PV panels is effective for maximising both SC and SS in households in Type 1 and 2, while increasing BESS capacity is more beneficial for households in Type 3 and 4.

\begin{figure}[!ht]
\centering
\includegraphics[width=0.45\textwidth]{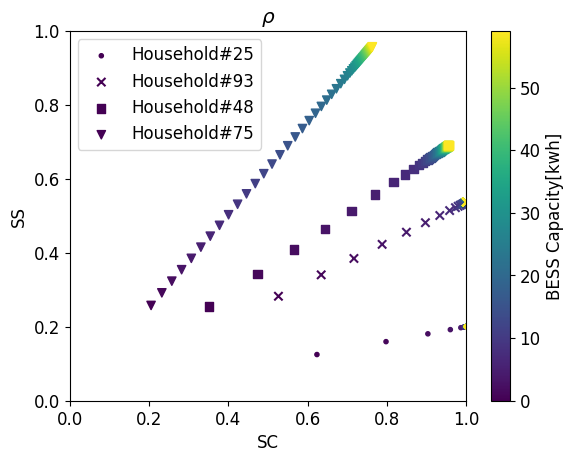}
\caption{$\rho$ for Different Household Types.}
\label{fig_relationship_rho}
\end{figure}

Fig.~\ref{fig_relationship_rho} shows the relationship factor between SC and SS in Households \#25, \#93, \#48 and \#75. As shown in Equation \ref{eq_rho}, the relationship between SC and SS is linear, where an increase in SC leads to a corresponding increase in SS. The $\rho$ values for Household\#25, \#93, \#48 and \#75 are 0.20, 0.54, 0.72 and 1.26, respectively. $\rho$ can indicate the maximum values of SC and SS regardless of the BESS optimisation. For example, Household \#25 shows that the maximum value of SS is about 0.2 while Household \#75 shows that the maximum value of SC is about 0.8, even as the battery capacity increases. Consequently, SC and SS should be analysed together rather than separately. A thorough understanding of their relationship allows households and operators to design PV-BESS systems more effectively.

\subsection{Optimal Operation Capacities}
\begin{figure}[!ht]
\centering
\includegraphics[width=0.45\textwidth]{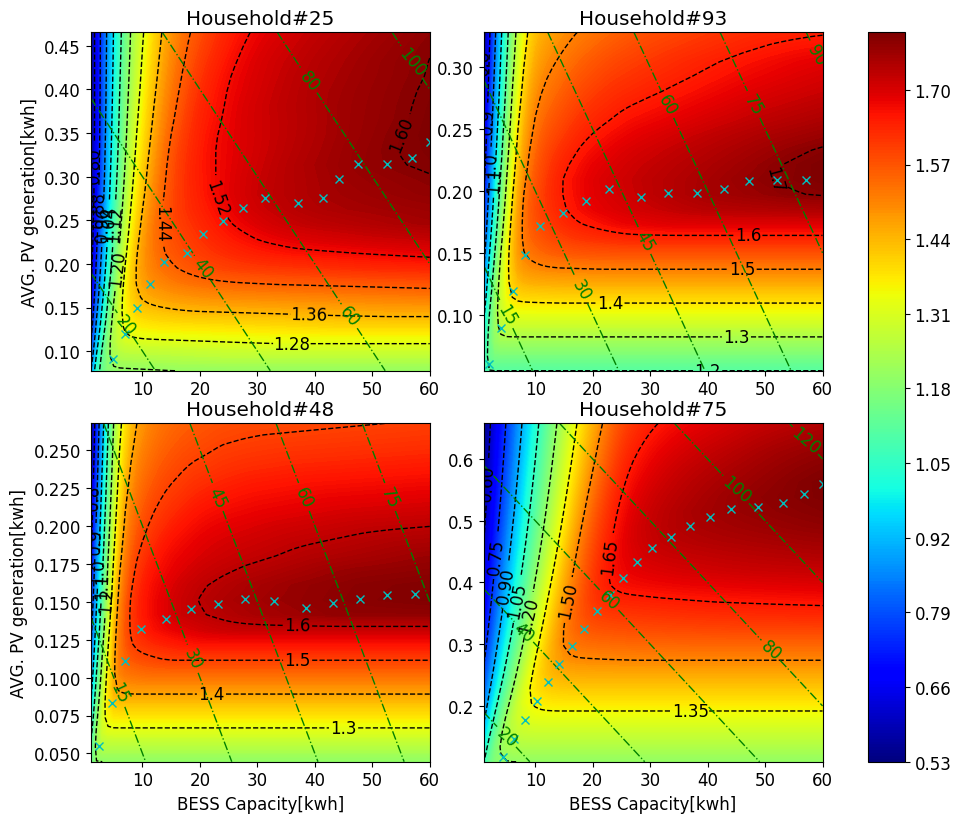}
\caption{Optimal Operation Points.}
\label{fig_optimal_points}
\end{figure}

To determine the optimal operation points for PV and BESS capacities in each category of household, we first define the cost of BESS capacity and the average generation capacity as follows:

\begin{equation}
    Cost_{res} = C_{batt} * SOC_{size} + C_{gen} * Avg(P_{gen})
\end{equation}
Where $C_{batt} and C_{gen}$ represent the cost of battery size per kWh and the cost of the average generation power per kWh, respectively. Fig.~\ref{fig_optimal_points} shows $SC+SS$ for each point of BESS and average generation power capacity in the case of $C_{batt}=1$ and $C_{gen}=100$. The color represents the level of $SC+SS$ and the green line indicates $Cost_{res}$. To maximise $SC + SS$, the optimal operation points of BESS and average generation power capacity, based on $Cost_{res}$, are mathematically calculated and shown as cyan cross marks.

\section{System Description}
In this section, the system model for MPC and RL models along with its objective function is discussed. Additionally, the model parameters used for the simulation are presented.

\subsection{System Model}

Each residential home has a PV solar and a BESS and the objective is to maximise $SC + SS$. The BESS management system is designed as follows:
\begin{equation}
\begin{split}
    P_{grid}(t) = &P_{load}(t) - P_{gen}(t)\\
    &+ P_{charging}(t) - P_{discharging}(t)
\end{split}
\end{equation}

\begin{align}
    \text{objective : } &max (SC + SS) \\
    \text{s.t. } & P_{charging}(t)\,P_{discharging}(t) = 0 \\
\begin{split}
    & SOC_{state}(t+1) = SOC_{state}(t) \\
    & \;\;\;\;\;\;\; +  P_{charging}(t) - P_{discharging}(t) \\
\end{split} \\
    & SOC_{min} \leq SOC_{state}(t) \leq SOC_{max}
\end{align}

This BESS operation model is used for simulating MPC and RL models.

\subsection{Model Parameters}
In this paper, the minimum and maximum SOC values are $10\%$ and $95\%$, respectively, and the initial SOC is $10\%$ of the battery capacity. Table \ref{tbl_params_bess_optimisation} shows hyper-parameters for each model.

\begin{table}[!ht]
\caption{Hyper-parameters of simulation models.}
\begin{tabular}{|l|c|c|c|c|}
\hline
                           & MPC & DQN      & A2C      & PPO      \\ \hline
Horizon                    & 48  & -        & -        & -        \\ \hline
LR                         & -   & 1.00E-05 & 1.00E-05 & 1.00E-05 \\ \hline
Batch size                 & -   & 128      & -        & -        \\ \hline
GAMMA                      & -   & 0.99     & 0.9      & 0.9      \\ \hline
Buffer size                & -   & 20000    & -        & -        \\ \hline
Value function coefficient & -   & -        & 0.5      & 0.5      \\ \hline
\end{tabular}
\label{tbl_params_bess_optimisation}
\end{table}

\section{Evaluation}
In this section, we present three optimisation scenarios: (i) BESS capacity optimisation, which optimises PV-BESS systems under various BESS capacities; (ii) PV-BESS optimisation, which optimises PV-BESS systems under varying BESS and PV capacities; (iii) PV-BESS optimisation with $\rho=1$, which optimises PV-BESS when $\rho=1$. The results from these scenarios are then compared. Note that these simulations are based on the real-world Ausgrid dataset.

\subsection{BESS Capacity Optimisation}
\begin{figure}[!ht]
\centering
\includegraphics[width=0.45\textwidth]{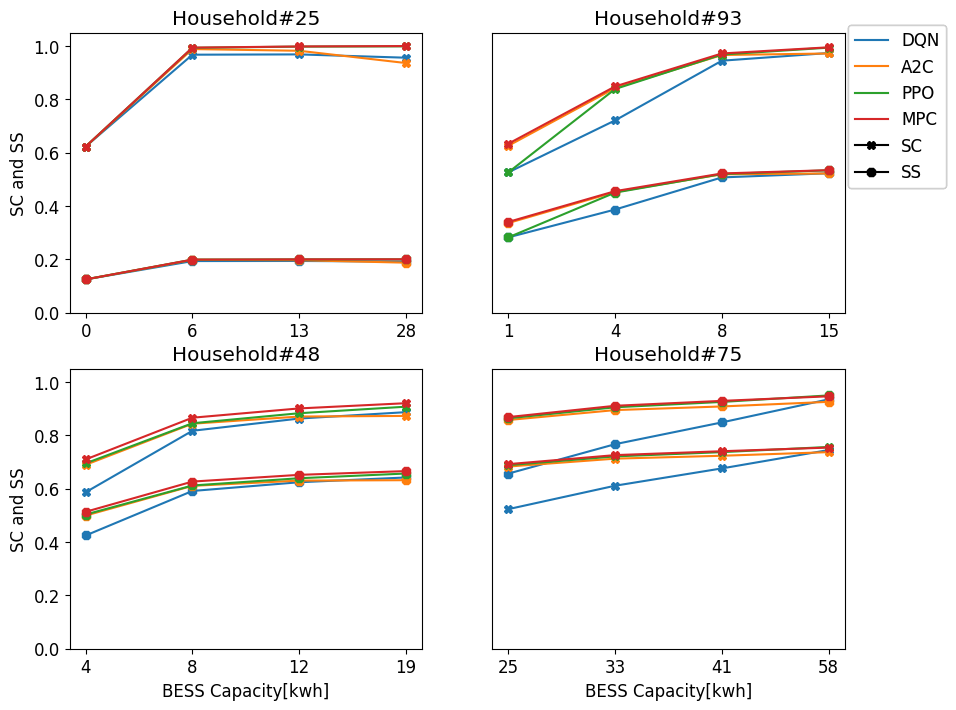}
\caption{BESS Capacity Optimisation Performance.}
\label{fig_bess_optimize}
\end{figure}

In this section, BESS capacity optimisation is performed by simulating the performance of MPC and RL models based on BESS capacities, without varying PV generation capacity. Fig.~\ref{fig_bess_optimize} illustrates SC and SS for each household, as simulated using MPC and RL solutions. MPC demonstrates the best performance, but RL solutions also perform similarly. In all cases, SC and SS do not improve even as BESS capacity increases. As shown in Table \ref{tbl_rho_bess_optimisation}, the $\rho$ values for each household are approximately the same as those derived by the LITE-BESS in the previous section.

\begin{table}[!ht]
    \centering
    \caption{$\rho$ values for BESS Capacity Optimisation.}
    \begin{tabular}{|l|l|l|l|l|}
    \hline
        ~ & DQN & A2C & PPO & MPC \\ \hline
        Household\#25 & 0.20 & 0.20 & 0.20 & 0.20 \\ \hline
        Household\#93 & 0.54 & 0.54 & 0.54 & 0.54 \\ \hline
        Household\#48 & 0.72 & 0.72 & 0.72 & 0.72 \\ \hline
        Household\#75 & 1.25 & 1.26 & 1.26 & 1.25 \\ \hline
    \end{tabular}
    \label{tbl_rho_bess_optimisation}
\end{table}

\subsection{PV-BESS Optimisation}
\begin{figure}[!ht]
\centering
\includegraphics[width=0.45\textwidth]{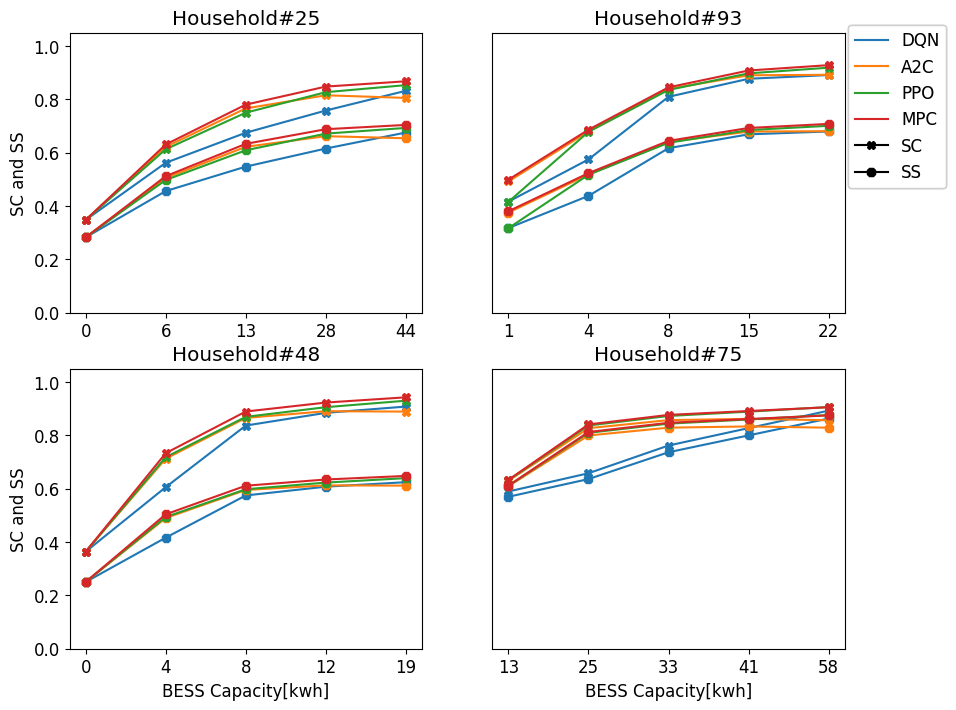}
\caption{PV-BESS Optimisation Performance.}
\label{fig_bess_pv_optimize}
\end{figure}

In this section, PV-BESS optimisation is performed by simulating the performance of MPC and RL models based on BESS capacities, while varying PV generation capacities, which are calculated using the mathematical method from the previous section. Table \ref{tbl_pv_capacity} presents the increase factors of the average PV generation power for each household, derived from the optimal operation points. The increase factors are applied to the PV generation power for each household to adjust the scale of PV generation without changing the patterns.  With PV-BESS optimisation, Fig.~\ref{fig_bess_pv_optimize} shows SC and SS for each household simulated using MPC and RL solutions using these increase factors in the Table \ref{tbl_pv_capacity}. The values of $\rho$ recalculated based on the varied increased factors and the $Cost_{res}$ indicate the total cost of PV and BESS capacities at the given BESS capacities. The average PV generation power increases for Household\#25 and \#93, while it decreases for Household\#75. 

\begin{table}[!ht]
\centering
\caption{PV-BESS optimisation Parameters.}
\begin{tabular}{|c|c|c|c|c|}
\hline
Household & $Cost_{res}$ (BESS Capacity) & Increase factor & $\rho$   \\ \hline
25        &  83 (51.5[kwh]) & x 4.05          & 0.81     \\ \hline
93        &  71 (50.1[kwh]) & x 1.42          & 0.77     \\ \hline
48        &  65 (49.7[kwh])& x 0.95          & 0.69     \\ \hline
75        & 106 (53.2[kwh])& x 0.77          & 0.96     \\ \hline
\end{tabular}
\label{tbl_pv_capacity}
\end{table}

As shown in Table \ref{tbl_rho_bess_pv_optimize}, the values of $\rho$ from the simulation results are approximately the same as the values of $\rho$ derived from the mathematical method.

\begin{table}[!ht]
    \centering
    \caption{$\rho$ values for PV-BESS Optimisation.}
    \begin{tabular}{|l|l|l|l|l|}
    \hline
        ~ & DQN & A2C & PPO & MPC \\ \hline
        Household\#25 & 0.81 & 0.81 & 0.81 & 0.81 \\ \hline
        Household\#93 & 0.76 & 0.76 & 0.76 & 0.76 \\ \hline
        Household\#48 & 0.69 & 0.69 & 0.69 & 0.69 \\ \hline
        Household\#75 & 0.97 & 0.97 & 0.97 & 0.97 \\ \hline
    \end{tabular}
    \label{tbl_rho_bess_pv_optimize}
\end{table}

\subsection{PV-BESS Optimisation with $\rho = 1$}

\begin{table}[!ht]
\centering
\caption{PV-BESS Optimisation with $\rho = 1$.}
\begin{tabular}{|c|c|c|c|}
\hline
Household & Increase factor & $\rho$  \\ \hline
25        & x 4.99          & 1.0     \\ \hline
93        & x 1.86          & 1.0     \\ \hline
48        & x 1.38          & 1.0     \\ \hline
75        & x 0.80          & 1.0     \\ \hline
\end{tabular}
\caption{PV-BESS optimisation Parameters with $\rho = 1$.}
\label{tbl_pv_capacity_rho}
\end{table}

\begin{figure}[!ht]
\centering
\includegraphics[width=0.45\textwidth]{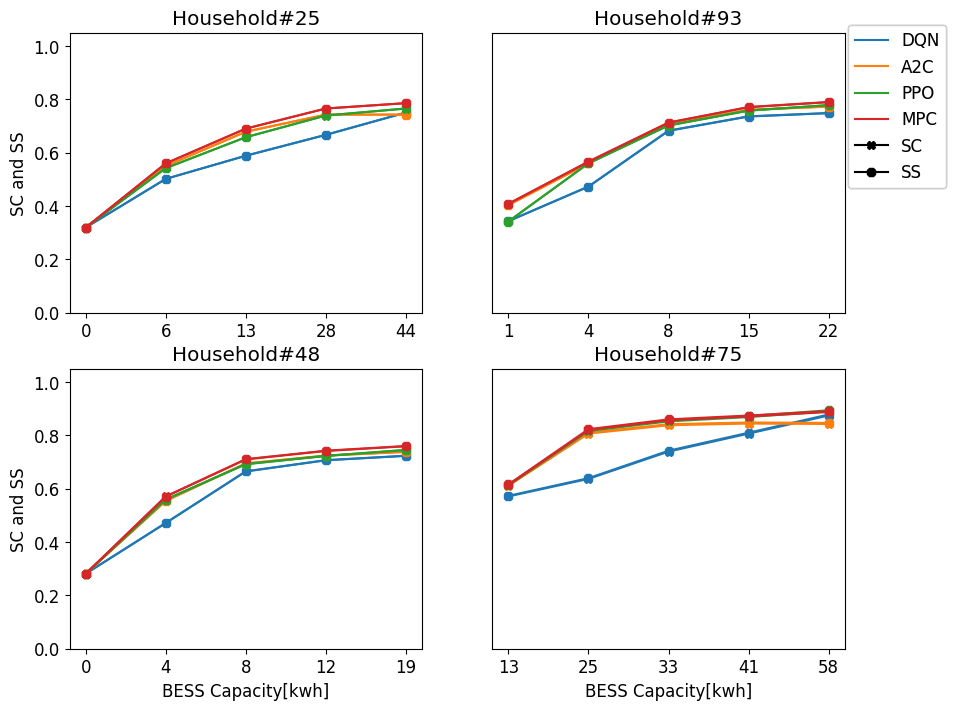}
\caption{PV-BESS Optimisation Performance with $\rho=1$.}
\label{fig_bess_pv_optimize_rho}
\end{figure}

In this section, PV-BESS optimisation with $\rho = 1$ is performed by simulating the performance of MPC and RL models based on BESS capacities, while varying PV generation capacities for $\rho = 1$. Table \ref{tbl_pv_capacity_rho} shows increase factors of PV generation capacities for $\rho = 1$ and Fig.~\ref{fig_bess_pv_optimize_rho} shows SC and SS for each household simulated using MPC and RL solutions with these optimisation parameters. In all cases, $\rho = 1$ is observed for each household, as expected. On the other hand, the levels of SS and SC vary when BESS capacity is 0 [kwh]. Additionally, the saturation BESS capacity and level for each household also differ. These differences arise from variation in power generation and consumption profiles and require further study.

\subsection{Performance Comparison}
\begin{figure*}[!b]
\centering
\includegraphics[width=0.9\textwidth]{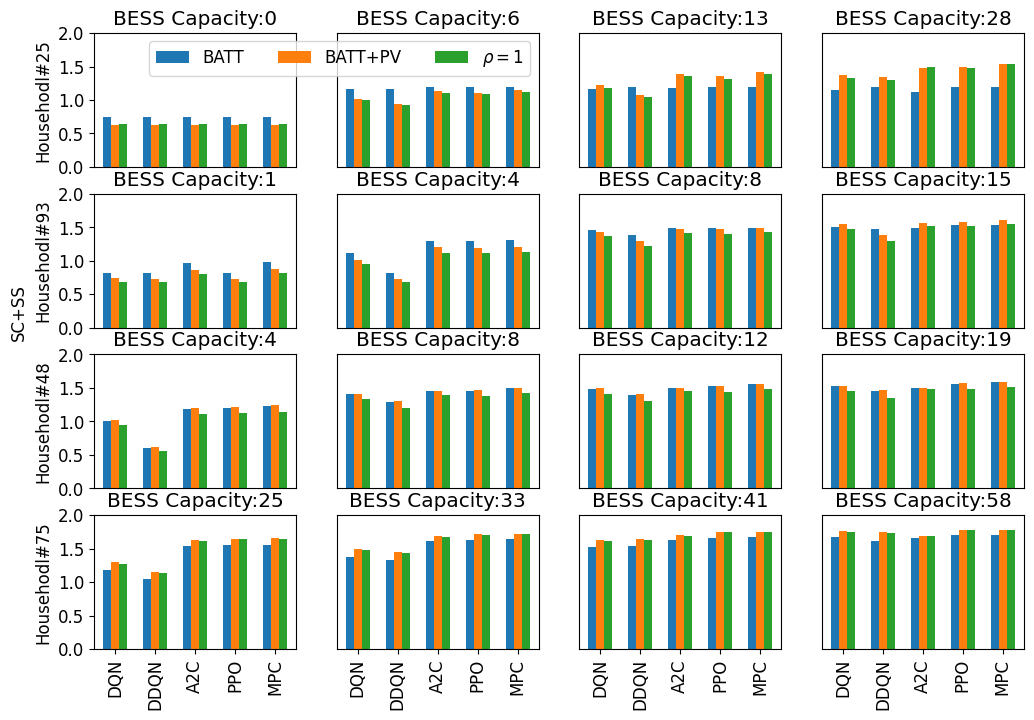}
\caption{Performance Comparison of PV-BESS Optimisations.}
\label{fig_compare_optimization}
\end{figure*}

Fig.\ref{fig_compare_optimization} shows the performance comparison between BESS optimisation, PV-BESS optimisation and PV-BESS optimisation with $\rho = 1$ for each household. When the average PV generation increases for household\#25 and \#93, $SC + SS$ increases as BESS capacity increases. In contrast, for household\#75, $SC + SS$ increases even though the average PV generation decreases. For household \#25 and \#75, the performance of PV-BESS and PV-BESS with $\rho = 1$ is similar, while for household \#93 and \#48, the performance of PV-BESS is better than that of PV-BESS with $\rho = 1$.

\section{Conclusion}
This paper presents an optimal operation strategy, LITE-BESS in a single household, focusing on the relationship between SC and SS. We mathematically define a relationship factor between SC and SS and analyse it using Ausgrid data, which includes PV generation and consumption of households with PV-BESS systems. Households are categorised into four groups based on this relationship factor using a simple and ultra-fast BESS management model. To evaluate these groups, we implement various BESS management models, including MPC and multiple RL approaches. Furthermore, we propose a method using the relationship factor to determine the optimal PV-BESS capacity for a single household and validate it through simulations with MPC and RL models. The result indicates that the relationship factor is useful in determining the operation capacity of PV and BESS to maximise $SC+SS$. The proposed solution is expected to assist households and operators in determining the appropriate capacity of PV and BESS when installing these systems at the household level.

\section*{Acknowledgments}
This work was supported by the Energy Transition Centre at Queensland University of Technology.



 
%

\bibliographystyle{IEEEtran}
\bibliography{References.bib}


\section{Biography Section}
\vspace{-1.0cm}

\begin{IEEEbiographynophoto}{Jun Wook Heo}
received the PhD degree in computer science from Queensland University of Technology (QUT), Australia. His research interests include distributed and decentralised solutions for large-scale networks, such as blockchain and zero-knowledge proofs.
\end{IEEEbiographynophoto}

\vspace{-1.0cm}
\begin{IEEEbiographynophoto}{Raja Jurdak (M'00-SM'11)} 
 received the BE degree in computer and communications engineering from the American University of Beirut, in 2000, the MS degree in computer networks and distributed computing from the Electrical and Computer Engineering Department, University of California, Irvine, in 2001, and the PhD degree in information and computer science, University of California, Irvine, in 2005. He is a Professor of Distributed Systems and Chair in Applied Data Sciences at Queensland University of Technology, and Director of the Trusted Networks Lab. His research interests include trust, mobility and energy-efficiency in networks. His publications have attracted over 10000 citations, with an h-index of 45. He serves on the editorial board of Ad Hoc Networks, Nature Scientific Reports, and on the organising and technical program committees of top international conferences, including Percom, ICBC, IPSN, WoWMoM, and ICDCS. He was TPC co-chair of ICBC in 2021. He is a conjoint professor with the University of New South Wales, and a senior member of the IEEE and a Distinguished Visitor of the IEEE Computer Society.
\end{IEEEbiographynophoto}
\vspace{-1.0cm}

\begin{IEEEbiographynophoto}{Sara Khalifa}
 received the PhD degree in computer science and engineering from the University of New South Wales, Sydney, NSW, Australia. She is an associate professor at Queensland University of Technology. Her research revolves around the broad aspects of ubiquitous sensing and edge computing for Internet of Things (IoT) applications. Specifically, she focuses on enhancing the energy efficiency of mobile sensing systems and developing  lightweight machine learning techniques for resource-constrained sensing devices. From 2016-2023, she was with CSIRO’s Data61 establishing the foundational research area of “Energy Harvesting Sensing (EHS)” as a core research focus developing a new paradigm for energy-efficient sensing and context recognition, opening up a multitude of new applications, generating IP, and attracting significant funding and commercial interest.
\end{IEEEbiographynophoto}

\vfill

\end{document}